\documentclass[journal,twoside,web]{ieeecolor}
\usepackage{generic}
\usepackage{cite}
\usepackage{amsmath,amssymb,amsfonts}
\usepackage{algorithmic}
\usepackage{graphicx}
\usepackage{textcomp}
\usepackage{multirow}
\def\BibTeX{{\rm B\kern-.05em{\sc i\kern-.025em b}\kern-.08em
    T\kern-.1667em\lower.7ex\hbox{E}\kern-.125emX}}
\begin{document}
\title{Exploiting full Resolution Feature Context for Liver Tumor and Vessel Segmentation via Integrate Framework: Application to Liver Tumor and Vessel 3D Reconstruction under embedded microprocessor}
\author{Xiangyu Meng, Xudong Zhang ,Gan Wang, Ying Zhang, Xin Shi, Huanhuan Dai, Zixuan Wang, and Xun Wang
\thanks{This research was supported by the National Natural Science Foundation of China [No.61972416, 61873280 and No.61873281]; the Natural Science Foundation of Shandong Province [No.ZR2019MF012]. (Xiangyu Meng and Xudong Zhang contributed equally to this work)(Corresponding author: Zixuan Wang and Xun Wang). }
\thanks{Xiangyu Meng, Xudong Zhang, Gan Wang, Ying Zhang, Xin Shi and Huanhuan Dai are with the Department of Computer Science and technology, China University of Petroleum, Qingdao, 266580, Shangdong, China(e-mail: xiangyumeng@s.upc.edu.cn; bigdongsir@163.com; s20070048@s.upc.edu.cn; zhangy9808@163.com; shix1104@163.com; daihuanhuan0901@163.com). }
\thanks{Zixuan Wang is with the Minimally Invasive Interventional Therapy Center, Qingdao Municipal Hospital, Qingdao, 266011, Shangdong, China(e-mail: prince\_room@sina.com).}
\thanks{Xun Wang is with the Department of Computer Science and technology, China University of Petroleum, Qingdao, 266580, Shangdong, China;  High Performance Computer Research Center, University of Chinese Academy of Sciences, Beijing, 100190, China(e-mail: wangsyun@upc.edu.cn).}}

\maketitle

\begin{abstract}
Liver cancer is one of the most common malignant diseases in the world.
	Segmentation and labeling of liver tumors and blood vessels in CT images can provide convenience for doctors in liver tumor diagnosis and surgical intervention. 
	In the past decades, many state-of-the-art medical image segmentation algorithms appeared during this period. 
	With the development of embedded devices, embedded deployment for medical segmentation and automatic reconstruction brings prospects for future automated surgical tasks.
	Yet, most of the existing segmentation methods mostly care about the spatial feature context and have a perception defect in the semantic relevance of medical images, which significantly affects the segmentation accuracy of liver tumors and blood vessels.
	Deploying large and complex models into embedded devices requires a reasonable trade-off between model accuracy, reasoning speed and model capacity.
    Given these problems, we introduce a multi-scale feature fusion network called TransFusionNet based on Transformer. 
    This network achieved very competitive performance for liver vessel and liver tumor segmentation tasks, meanwhile it can improve the recognition of morphologic margins of liver tumors by exploiting the global information of CT images.
	Experiments show that in vessel segmentation task TransFusionNet achieved mean Dice coefficients of 0.899 and in liver tumor segmentation task TransFusionNet achieved mean Dice coefficients of 0.961. 
	Compared with the state-of-the-art framework, our model achieves the best segmentation result.
In addition, we deployed the model into an embedded micro-structure and constructed an integrated model for liver tumor vascular segmentation and reconstruction. 
This proprietary structure will be the exclusive component of the future medical field.
\end{abstract}

\begin{IEEEkeywords}
Liver tumor, Liver vessel, Medical image segmentation, Transformer, 3D reconstruction,  embedded microprocessor, computer-aided diagnosis
\end{IEEEkeywords}

\section{Introduction}
\label{sec:introduction}
\IEEEPARstart{L}{iver} cancer is the sixth most common primary cancer worldwide and the fourth leading cause of cancer death\cite{li2021immunological}. 
Therefore, there is an urgent need for effective prevention programs and treatments to reduce the harm caused by liver cancer. 
In the early stage of liver cancer, potential risks of coming serious liver cancer can be eliminated by surgical removal of the tumor or local treatment. 
In recent years, computer-assisted liver surgery (e.g., ablation and embolization) has been increasingly used for the treatment of primary and secondary liver tumor patients who are not eligible for common surgeries\cite{gervais2009society}. 
Computed Tomography(CT), as part of computer-assisted liver surgery, is a commonly implemented for clinical diagnostic approach to improve the visualization on liver, vessels and tumors\cite{ciecholewski2021computational}. 
Because CT is capable of clearly showing the number, boundary, density and other patterns of the disease focus. 
Experts will segment liver vessels and tumors from CT images before surgery in helping 3D visualization, path planning, and guidance for interventional surgery of liver\cite{yan2020attention}.
However, there are some challenging obstacles in computer-assisted liver interventions. 
The most critical one of all is that segmentation of liver vessels and tumors from CT images is manually completed by specialists, which is rather time-consuming, labor-intensive and no quality guaranteed. 
This can lead to the inability to precisely pinpoint the vessels that supplies nutrition for the hepatic tumor, thus affecting hepatic embolization procedure, ablation and so on. 
Eventually local tumors will relapse\cite{huang2013influence} . 
As a result, there is an urgent need for an intelligent auxiliary diagnostic key embedded component in the medical field. The structure can be flexibly deployed in any CT instrument. Meanwhile inferential results of liver tumor and artery can be quickly generated with guaranteed precision, which assist physicians to complete rapid diagnosis and carry out next liver surgery plan.

In previous studies, many methods have emerged for segmenting liver vessels separately or tumors as well, but none of which considers segmenting vessels and tumors at the same time. This is due to the complicated background, heterogeneous shape and surrounding vessels irregularity of the tumor making it difficult to segment the hepatic vessels that supply nutrition for the tumor\cite{jiang2019ahcnet}. 
Traditional methods try to segment livers or tumors by active contour methods, tracking methods, and feature learning methods. 
Active Contour Model (ACM) is a method to detect object boundaries based on curve evolution theory and level set approach. Cheng et al. \cite{cheng2015accurate} 
implemented ACM with precise shape dimension constraints based on CT scan models for contour point detection of vessel cross-sections to plot vessel boundaries. 
Chung et al. \cite{chung2018accurate} proposed an active contour method to segment portal vein and hepatic vein based on the regional intensity distribution of the image and the probability map of vessel occurrence.
However, the active contour model tends to fall into the local optimum problem when extracting complex regions in the vector field, and cannot handle gray scale inhomogeneous images well. 
The tracing method starts by manual initialization or image preprocessing to initiate a single or specified number of seed points in the vessel, and then finds subsequent points based on the image derived data as a way to trace the vessel\cite{ciecholewski2021computational}. 
Tracking methods mainly include model-based algorithms \cite{bauer2010segmentation,esneault2009liver,lebre2019automatic},least cost path-based algorithms \cite{kaftan2009two,zeng2017liver}.
However, if the initial seed points of these methods are not correctly positioned, the final segmentation results can be seriously affected.

In order to segment vessels or tumors from CT images, feature learning methods need to perform feature extraction from images and labels based on real segmentation to train machine learning models such as random forest (RF) \cite{mahapatra2014analyzing,smith2010image} and support vector machine (SVM) \cite{wanga2011color,zhiwen2011modified} for automatically segmenting vessels or tumors from CT images.
However, the robustness and generalization ability of machine learning models are limited. 
In recent years, many deep learning models ,like convolutional neural networks \cite{badrinarayanan2017segnet,meng2021computationally},have gradually shown promosing performances in the field of medical image segmentation. 
Currently, segmentation models based on fully convolutional networks \cite{long2015fully} and UNet \cite{ronneberger2015u} architectures are the most effective ones. 
Huang et al. \cite{huang2018robust} combined 3D-UNet with data enhancement techniques, a variant of dice coefficient, to reduce the effect of high imbalance in some extent between hepatic vessels and background classes. 
Zhou et al. \cite{zhou2018unet++} proposed UNet++, a model that combines a deeply supervised encoder and decoder and links the sub-networks of both through a series of hops as a way to reduce the semantic gap between the encoder and decoder feature mappings. 
Recently, Transformer \cite{vaswani2017attention}\cite{0AMDE}\cite{SONG2022} has made great achievements in the field of deep learning, and TransUNet proposed by Chen et al. applies transformer as an encoder to extract global contextual features and combines it with convolutional neural network for decoding. 
For segmentation of liver vessels and tumors, a high degree of accuracy must be achieved to enable clinical applications. 
In view of above mentioned methods including other UNet-based \cite{li2020anu,jha2020doubleu,song2019u} methods , performances still can be improved in terms of accuracy and efficiency despite of some attempts in architecture.

Because embedded microprocessors are low-power, inexpensive, and easy to deploy, researchers considered migrating semantic segmentation models to embedded microprocessors or edge computing devices in order to complete the inference task in specific scenarios. 
Wei et al.\cite{2019ThunderNet} introduced a fast and efficient lightweight network called Turbo Unified Network (ThunderNet).
This model implements fast and efficient inference on the Jetson platform.
Huang et al. demonstrate EDSSA-an Encoder-Decoder semantic segmentation networks accelerator architecture which can be implemented with flexible parameter configurations and hardware resources on the FPGA platforms that support Open Computing Language (OpenCL) development\cite{2020EDSSA}.
In the process of model transplantation, the tradeoff between speed, volume and accuracy of model inference is the focus of various researchers.
With the development of medical image segmentation methods, related studies show significant precision in different lesion segmentation of multimodal medical data. 
Lightweight model deployment and transplantation of high-precision medical segmentation models into embedded micro-devices will greatly promote the development of automated surgery and automated diagnosis.

In our work, we construct a general semantic segmentation model TransFusionNet according to the different semantic features of liver tumors and vessels in CT images. 
It is a multiscale information fusion network capable of learning sematic and spatial information features, including a Transformer-based semantically feature extraction module, a Multi-layer local feature extraction module, and a multiscale fusion decoder. 
The model can accurately detect and segment fine-scale arterial vasculature, while effectively identifying and segmenting liver tumors and vascular features by fusing the global context of CT images. 
At the same time, based on the Edge Extraction Module, the network can effectively extract the edge features of the target objects.
Since images were obtained from different scanners or different imaging protocols during clinical diagnosis. 
Instead, training of deep learning models often requires a large amount of labeled data that is supposed to accurately represent the original data\cite{van2014transfer}.  
To address the above issues, we propose a transfer learning training strategy. This learning strategy allows the network to mix the features of the three datasets, and the final trained model can significantly alleviate overfitting and improve segmentation accuracy.
Finally, we deployed the model to the Jetson TX2 embedded microprocessor using compressed distillation to allow the model to segment CT medical images in real-time and rebuild tumors and blood vessels. 
Experiment shows that this inference system can complete fast and automated reconstructions with the error's permission, which can greatly reduce the workload of doctors in diagnosis. 
In the future, this structure can be used as a key component in intelligent surgical diagnosis and has high application prospects. 
At the same time, the portability of our proposed framework can be demonstrated by transplanting high-precision medical segmentation models.

\section{Methods}
\begin{figure*}[!t]
	\centering
	\centerline{\includegraphics[width=0.8\textwidth]{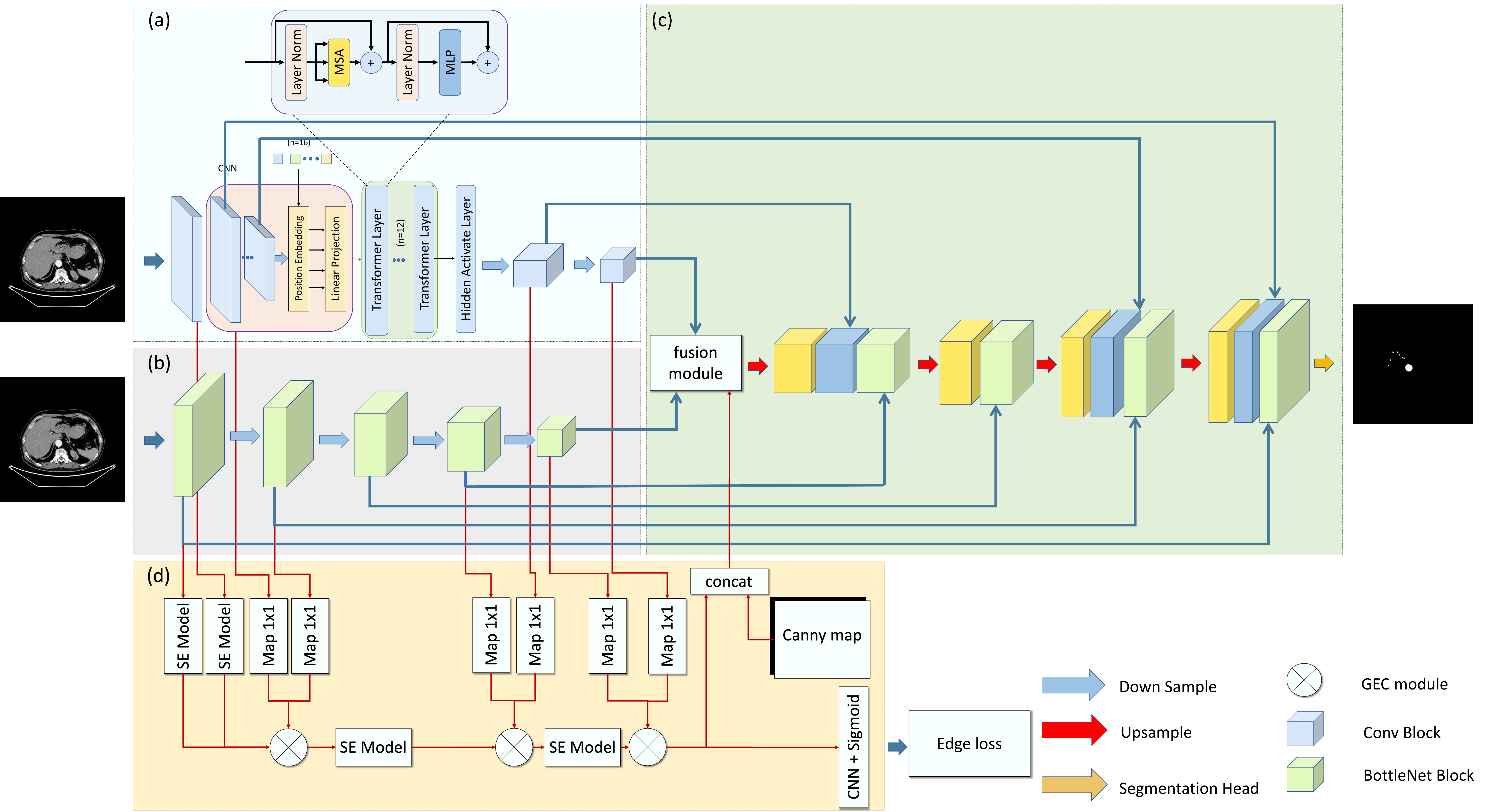}}
	\caption{Overview of the TransFusionNet model. (a) Transformer-based feature extractor. (b) Multi-layer local feature extraction module (c) Fusion decoder for multiscale feature. (d) Edge Extraction Module.}
	\label{fig:fig1}
\end{figure*}

In order to better complete the segmentation task of liver tumors and blood vessels, we design a novel segmentation architecture called TransFusionNet. We introduced a new feature extraction module fused with Transformer and CNN. Based on this module, the network can effectively extract image spatial features and semantically related features. At the same time, we proposed an Edge Extraction Module which can significantly capture the edge feature of the image to cooperate with the training of the segmentation network. The model designed based on our ideas can effectively learn rich feature information, and can effectively ensure the segmentation accuracy of the edges of difficult-to-segment objects, which are critical in vascular and liver tumor segmentation tasks. The framework of our model is shown in Figure \ref{fig:fig1}.

\subsection{Transformer-based semantically feature extraction module}

We introduce an encoder that can learn the global feature representation, which consists of a feature embedding module based on a feature extraction backbone and a feature extraction module that senses the semantically related information representation of the image based on the transformer \cite{vaswani2017attention}. 
This module adopts a brand-new feature extraction idea, by semantically representing the features of the picture and learning the global representation of semantic features.

The input image $i \in \mathbb{R}^{H \times W \times C}$  is first fed into the feature extraction backbone network. 
The network can extract the spatial information features of CT images and output the feature map $x \in \mathbb{R}^{H^{'} \times W^{'} \times C^{'}}$.
We divide the feature map x learned by the backbone into a series of patches  $x_p^i \in \mathbb{R}^{P^2 \times C},i=1,…,N$, where the size of each patch is $P \times P$,and the number of patches denote by $N=\frac{H^{'} \times W^{'}}{P^2}$. 
For each patch, we use a convolution operation with a kernel size of $P \times P$ to obtain the information $E_{info}^i$ of i-th patch to form an information matrix $\{E_{info}^1,E_{info}^2,...,E_{info}^N\}$.
In order to better learn location information using Transformer, Dosovitskiy et al. \cite{dosovitskiy2020image} perform a learnable location embedding for each patch to obtain the location matrix $\{E_{pos}^1,E_{pos}^1,...,E_{pos}^N\}$ of the N patches. 
The feature of the i-th patch can be formulated by the following equation:
\begin{equation}
    E^i=E_{info}^i+E_{pos}^i.
     \label{eq1}
\end{equation}
We adopt this position encoding method, so that the feature extraction module can effectively learn the position information of the features.
We next input the above obtained feature matrix $E=\{E^1,E^2,…,E^N \}$ of x into multi Transformer layers to learn semantically representation of the feature map. 
In comparison to traditional convolution operation, transformer adopts a multi-head self-attention mechanism, and its core formulation is shown in equation \ref{eq2}:
\begin{equation}
    y=\sum_{i=1}^h \sum_{j=1}^w \sum_{k=1}^n(softmax(Q_{ijk}^T \times K_{ijk}) \times V_{ijk}),
    \label{eq2}
\end{equation}
where $h$ and $w$ denote by width and height of the feature matrix $E$ after feature extraction and location embedding. 
And $n$ is the number of self-attention mechanism heads.
$Q_{ijk},K_{ijk},V_{ijk}$ denote the query, key and value obtained by three linear transformations of the input $E_{ij}$ in each self-attended head, respectively.
$y \in \mathbb{R}^{h×w×C}$ denotes the output after one multi-headed self-attention.
We stacked 12 transformer layers, and the output of the last layer can theoretically learn to incorporate a rich context feature representation of the CT image under a wider range of perceptual fields. 
We then feed the output of the Transformer layers into a three-layer convolution operation. 
The final output feature map consist global high-level abstract information, effectively solving the problem of missing information caused by perceptual field defects in traditional deep CNN networks.
We call it semantic feature map.

\subsection{Multi-layer local feature extraction module}

\begin{figure}[!t]
    \centering
    \centerline{\includegraphics[width=0.8\columnwidth]{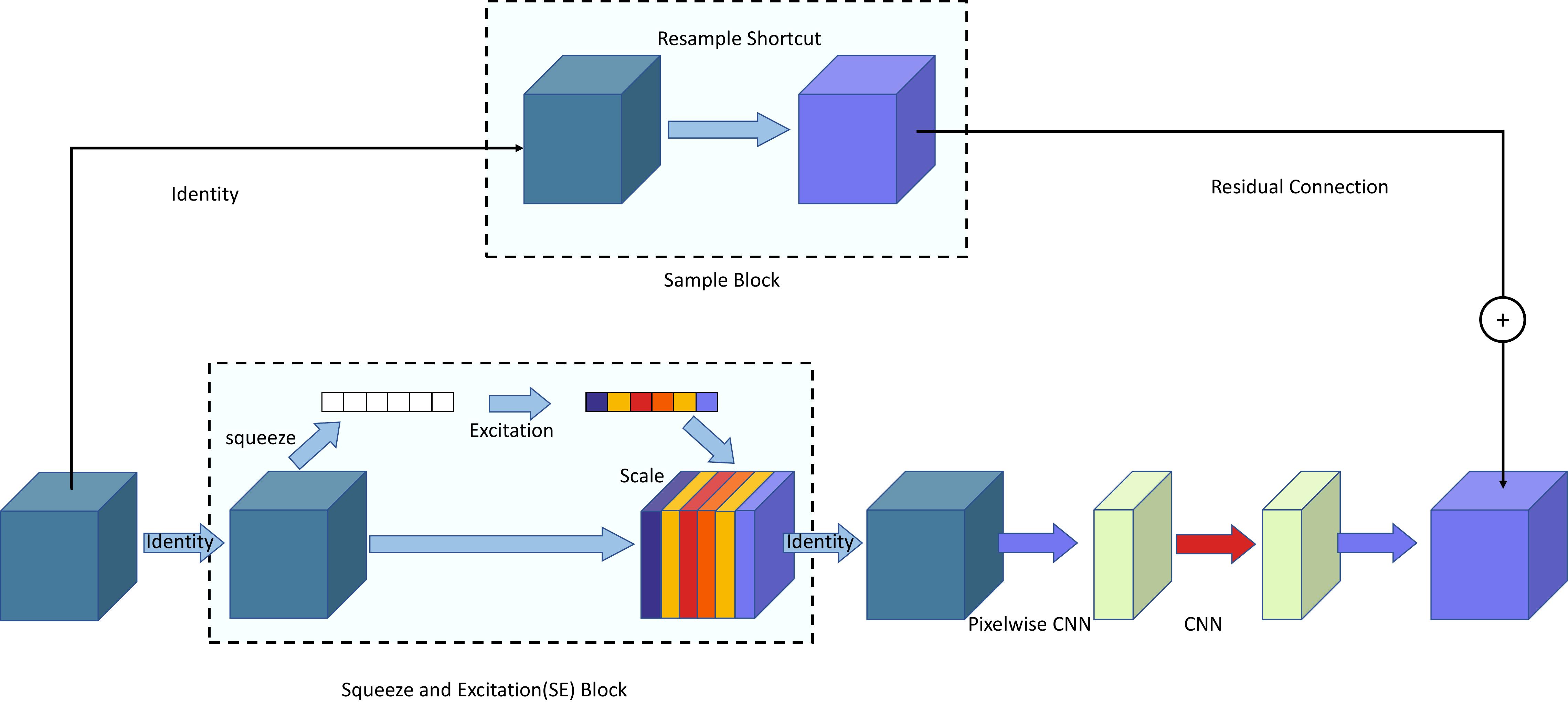}}
    \caption{BottleNet network structure with SEblock}
    \label{fig:fig2}
\end{figure}

Transformer-based extraction module is a very powerful for semantically information feature, because the Transformer feature extraction module has advantages in learning semantically related features.
In many ways, however, Transformer is not an effective replacement for traditional convolutional operations. 
For extraction of more subtle feature in some images such as edge feature of interest regions and tiny vessel feature, CNN is nothing but the perfect solution.
We designed a local residual network encoder based on multi-layer SEBottleNet stacking, as shown in Figure \ref{fig:fig2}. 
The encoder consists of a six feature extraction module.
A max pooling operation is performed to extract the high-level feature representation after feed feature map to each feature extraction block.
The input CT image $x \in \mathbb{R}^{H \times W}$ is first fed forward to a CNN module for high-level feature extraction, and the feature map $u\in R^{H \times W \times C}$ is obtained. 
Then, the feature map $u$ is feed into a deep residual feature extractor stacked by five layers of SEBottleNet, each of which is used for learning the context features under the local perception field.
BottleNet residual network\cite{he2016identity} retains all the advantages of residual network and significantly reduces computation interval and computational burden. 
We introduced the Squeeze and Excitation(SE)\cite{hu2018squeeze} in the BottleNet to enhance the interdependence between feature map channels.
The structure of the SEBottleNet is shown as in Figure \ref{fig:fig2}.
The mean value $e_c \in \mathbb{R}^{1 \times 1 \times C}$ of the feature embedding for each channel in the feature map $U \in \mathbb{R}^{H \times W \times C}$ can be obtained from the Squeeze section, as shown in the following equation:

\begin{equation}
    e_c=\frac{1}{w \times h} \sum_{i=1}^w \sum_{j=1}^h u_c(i,j).
    \label{eq3}
\end{equation}
Where the $u_c(i,j) \in \mathbb{R}^{1 \times 1 \times C}$ is the pixel in feature map $U$.
The Excitation section can learn the feature weights $e_c$ for each channel by $s_c$:

\begin{equation}
    s_c=\delta(\mathcal{G}(e_c,\mathcal{W})).
    \label{eq4}
\end{equation}
Finally, the vector product $\tilde{O}$ of s and u is obtained by the Scale operation, and this is the final output of the SE module:

\begin{equation}
    \tilde{O_c}=s_c \times u_c,
    \label{eq5}
\end{equation}
where $\tilde{O}_c$ is the feature map of a feature channel.


The SEBottleNet residual network splits the traditional convolutional operation into multiple modules to ensure that each module has a different feature extraction task. 
We introduced the Squeeze and Excitation module in the middle of the module to better learn the importance of the feature map channel dimensions, so that SEBottleNet has a stronger learning focus in the feature extraction process. 
Through the continuous stacking of SEBottleNet and maxpool, the encoder can continuously extract the local feature representation of the input CT image HIGH-level. 
Meanwhile, since each SEBottleNet is set with residual connections, it enables the encoder to effectively mitigate the degradation problem caused by network deepening.

\subsection{Edge Extraction Module}

Since the hepatic arterial vessels are very small and the margins of the liver tumor arblurd, further refining the segmentation of the vessels and liver is a challenging task. 
In order to allow the model to learn more detailed spatial feature features, we introduce the Edge Extraction Module, which is specially designed to learn the edge features of blood vessels and tumor regions of interest and fuse the edge features to the segmentation network. 
The structure of this module is shown in Figure \ref{fig:fig1} (d). 
The Edge Extraction Module takes the feature maps of feature extraction layers and the CT edge map(Figure \ref{fig:dataset}(b)) extracted by the Canny algorithm as input, and predict the edge result $e \in \mathbb{R}^{H \times W }$.
This module predicts edge information and fuses the predicted feature maps into the segmentation network. 
To accomplish this task, we process segmentation annotations to obtain edge annotations $e_r$(Figure \ref{fig:dataset}(d)), which can be used as a supervision condition for this module.

In this module, we used Gated Excitation Convolution (GEC) layer. 
GEC is the most important unit in Edge extraction module and it can filter out some irrelevant information to help Edge extraction module focus on extracting image edge features. 
GEC is applied between the Edge extraction module and the feature extraction module. 
It uses gating mechanisms to deactivate its own activations that are not deemed relevant by the higher-level information contained in the extraction module\cite{takikawa2019gated}. 
At the same time, we introduce an excitation module in the gating activation layer to learn the importance between different feature maps. 

We define $t_i,c_i \in \mathbb{R}^{\frac{H}{2i} \times \frac{W}{2i} \times C}$  as the feature maps of the Transformer module and the local feature extraction module, and $i$ denote the number of locations.
Before using the GEC module, $t_i$ and $c_i$ were fed into a convolutional layer $C_{1 \times 1}$ to obtain the image-dimensional feature maps $t_{i}^{'},c_{i}^{'} \in \mathbb{R}^{H \times W}$.
Let $e_i \in \mathbb{R}^{H \times W \times C}$ denote the feature map synthesized by the Edge Extraction Module.
Given the feature maps $t_{i}^{'},c_{i}^{'}$ and $e_i$,  an excitation convolutional layer is applied to generate the sigmoid activation $\alpha_{i} \in \mathbb{R}^{H \times W}$:
\begin{equation}
    \alpha_{i}=\sigma(C_{1 \times 1}(F_{se}(cat(t_{i}^{'},c_{i}^{'},e_i)))),
    \label{active}
\end{equation}
where $F_{se}$ denotes squeeze and excitation option as shown in Equation \ref{eq3}-\ref{eq5}. 
Finally, $\alpha_{i}$ and $e_i$ is fed into Gated Convolution layer\cite{Yu_2019_ICCV}\cite{dauphin2017language}\cite{takikawa2019gated} and generate the $e_{i}^{'}$. 
The Gated Convolution layer is compute as 
\begin{equation}
    e_{i}^{'} = C_{1 \times 1}((e_{i}^{'} \times \alpha_{i}) + e_{i}^{'}).
    \label{active}
\end{equation}
Theoretically, GEC can be simply regard as a collection of attention for the spatial dimension and channel dimension of the feature map. 
Through GEC operation, the attention maps $\alpha_{i}$ selectively preserve the edge semantic features.
We cancel the GEC operation on the shallow feature maps of the feature extractor, since the images feed to the convolution layer mainly learns the general low-level features, at the same time, the output feature map retains rich edge information. 
As the network deepens, the feature map will retain high-level features. 
Using GEC operation can effectively weight the useful edge information of high-level features in theory.

The Canny operator can effectively filter out the irrelevant features of the image to obtain the canny image as shown in Figure \ref{fig:dataset}(b). 
We think it is applicable for medical image segmentation. 
Therefore, we firstly concatenate the canny image and the last GEC module output $e_n$. 
Then we feed them together with the output feature maps of the two feature extractors to the Fusion Module. 
At the same time, the edge extraction module uses edge loss as the loss function and edge label as the supervision to optimize the prediction edge map.

\subsection{Multi-scale feature fusing module}

\begin{figure}[!t]
	\centering
	\centerline{\includegraphics[width=\columnwidth]{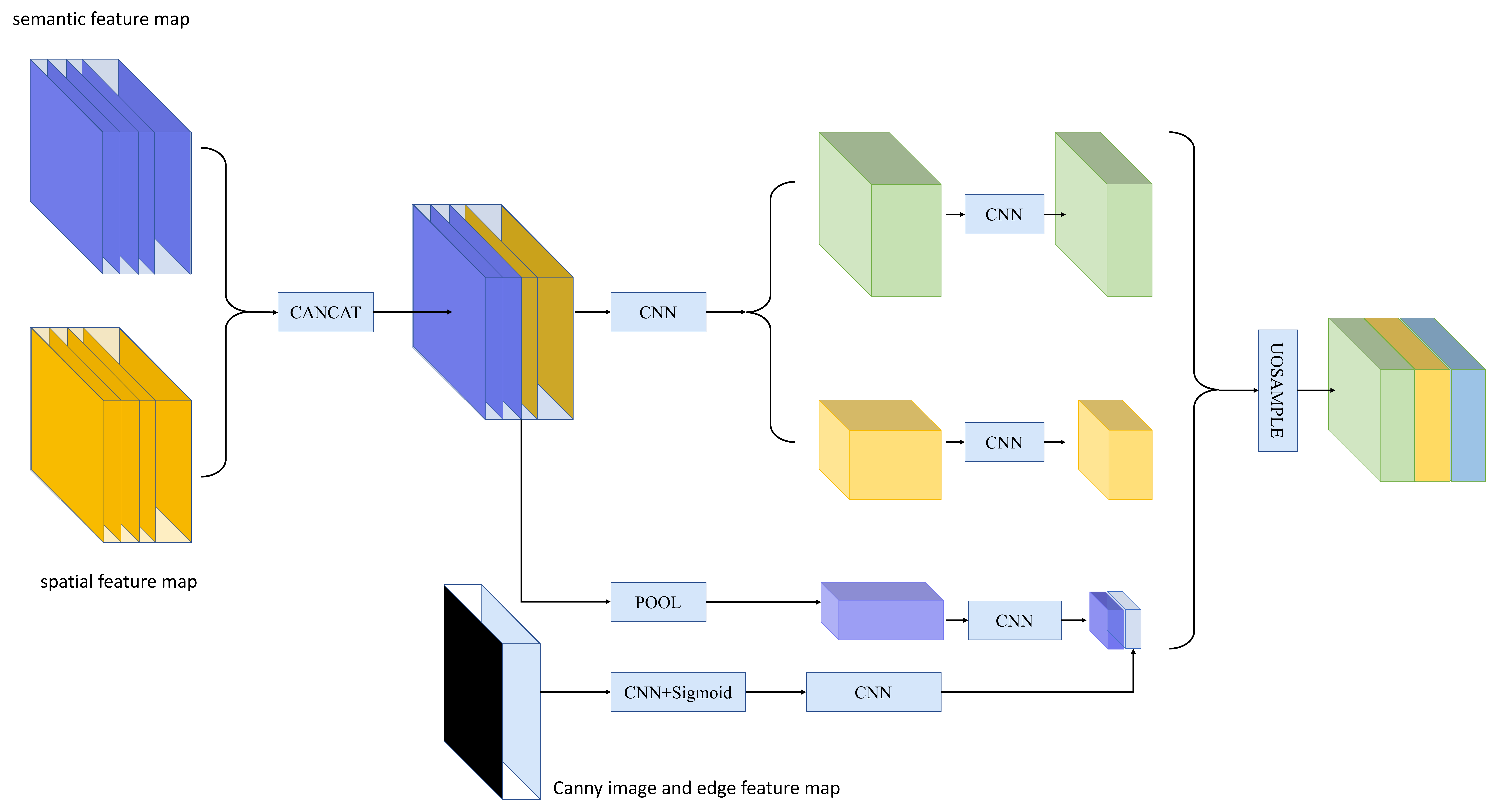}}
	\caption{Network framework of fusion module}
	\label{fusion}
\end{figure}

In the previous sections, we introduced two feature extraction structures that learn spatial features and semantically related features, respectively. 
At the same time, we introduce the edge feature extraction module to enhance the ability of the whole network to extract the target edge features.
In this section, we introduce the multi-scale feature fusion decoding module to sample the semantic features learned by the three modules.
This module takes the feature maps extracted by the three modules as the input and outputs the predicted category distribution map $\hat{y} \in \mathbb{R}^{K \times H \times W}$, where K represent the semantic classes.

We introduce a fusion module, which mainly fuses the feature maps of the three feature extraction modules. 
Figure \ref{fusion} shows the structure of this module.
We designed the module with reference to the spatial pyramid pooling(SPP). 
Firstly, the module uses $C_{1 \times 1}$ and $C_{3 \times 3}$ convolution to extract features from the concatenation result of the semantic feature map and spatial feature map respectively. 
Next, we feed it into the pooling layer and fused the edge feature map. 
Through the above operations, we obtained the feature maps of three different receptive fields. 
Finally, we sample and concatenate these three feature maps to output the fused feature map.
Theoretically, the feature map output by these module can retain rich spatial features, semantic related features and edge features.

In the process of continuous feature extraction layer by layer in the coded network, the low-level information of the feature map is continuously filtered and the high-level information is extracted. 
UNet uses skip connections to conduct the feature maps of the encoding module of each stage to the decoding module of the corresponding stage, and the network can fully learn the feature maps of different levels of the image. 
We adopted the skip connection operation from UNet and introduced skip connections to different feature encoders to allow the whole network to better learn the feature information of different encoders at different levels. 
The skip connection introduced in the local feature extraction module is similar to the traditional UNet module, which combines the short-range skip connection (residual connection) and the long-range skip connection of SEBottleNet. 
As for the Transformer based feature extraction module, we first introduce skip connections in the encoding process of the backbone network to connect the intermediate feature maps in the forward propagation process of the backbone embedding network, which improves the low-level feature loss in the feature embedding process of the backbone network. 
Next, we add skip connections to the feature maps with global feature representations after Transformer feature encoding fusion to fuse the global low-level features. 
Eventually, after continuously fusing low-level feature maps of different scales, the decoder can learn the semantic information of images from coarse to fine.

\subsection{multi task training strategy}
We propose the edge feature extraction module to cooperate with the segmentation task of the model, so we train the model to complete semantic segmentation and edge information segmentation at the same time. 
We introduce the joint optimization of edge loss and segmentation loss respectively. 
At the same time, in order to better ensure the consistency of multi task learning optimization, we set up a regularization methods to balance the two losses.

We use Dice and Cross Entropy(CE) as the loss function of segmentation task to predict semantic segmentation $y$:

\begin{equation}
    \mathcal{L}_{seg}^{\theta, \psi} =\lambda_{1} \mathcal{L}_{Dice}(y,\hat{y})+ \lambda_{2} \mathcal{L}_{CE}(y,\hat{y}),
    \label{seg_eq}
\end{equation}
where $y \in \mathbb{R}^{H \times W}$ denote the real semantic label map of liver tumor and vessel.
In the equation \ref{seg_eq}, $\lambda_{1}$ and $\lambda_{2}$ represent hyper parameters. 
As for edge prediction, we use Binary Cross Entropy(BCE) loss. 
In this experiment, the model mainly focuses on tumor and vascular segmentation. 
We extract their common edges to obtain edge label $\hat{e} \in \mathbb{R}^{H \times W}$ (Figure \ref{fig:dataset}(d)) and take $\hat{e}$ as the loss supervision.
Therefore, the edge loss can be expressed as:
\begin{equation}
    \mathcal{L}_{edge}^{\theta, \psi} = \lambda_{3} \mathcal{L}_{BCE}(e,\hat{e}),
    \label{edge_eq}
\end{equation}
where, $e$ represent the edge predict map of the edge extraction module.
It is worth noting that in the optimization process, the parameters of feature extraction modules and edge extraction modules will be optimized based on loss. 
Next, we will input the feature map output by the edge extraction module into the fusion module to predict the segmentation results. 
Therefore, the prior knowledge learned by the edge extraction module was retained in $y$. 
At the same time, the segmentation loss will pay more attention to the edge features in the optimization process.

We introduce regularization methods to make the model cooperate better in the training process. 
As mentioned above, $y \in \mathbb{R}^{K \times H \times W}$ represents the predicted segmentation map and $e \in \mathbb{R}^{H \times W}$ represents the predicted edge graph. 
Therefore, we introduce shape regularization, which can be expressed as:
\begin{equation}
    \mathcal{L}_{sreg}^{\theta, \psi} = \lambda_{4} ||Sigmoid(\oplus y) \times e - e||,
    \label{sreg_eq}
\end{equation}
Where $\oplus$ represents the pixel-wise addition of $y$ without the background label map which can be implemented using kernel fixed convolution operator. 
This operation outputs a label map containing the predict region of the tumor and blood vessels.
In particular, at the beginning of model training, because the edge extraction module cannot accurately predict the edge, loss does not play any role. 
We therefore introduced a  dynamic adjustment strategy, which is set $\lambda_{4}$ to 0 before 100 epoch and $\lambda_{4} \geq 0$ after 100 epoch

Finally, the loss function of the model is:
\begin{equation}
    \mathcal{L}_{total} = \mathcal{L}_{seg}^{\theta, \psi} + \mathcal{L}_{edge}^{\theta, \psi} + \mathcal{L}_{sreg}^{\theta, \psi}.
    \label{total_eq}
\end{equation}
We set the epoch to 300, the initial learning rate to 0.001 (using the cosine annealing learning rate decay method), and the batch size to 8. 
The model is trained using an SGD optimizer with a momentum of 0.9 and a weight decay of 1e-4.
\subsection{Applying transfer learning to TransFusionNet}

The TransFusionNet can significantly learn full-resolution context feature information, and its segmentation effect in the public dataset of blood vessels and liver tumors is significant. 
However, due to the scarcity of the enhanced CT images of liver cancer after the screening, we only obtained CT images of 18 patients. 
Too little data will inevitably affect the performance of the model and deepen the over-fitting problem.
For this purpose we introduce a transfer learning strategy, which does not require exactly representative training data and is able to take advantage of the similarity between datasets to capture specific prior knowledge during the training phase of the model in order to construct new segmentation models.

We first pre-trained the models using the publicly available datasets LITS and 3Dircadb to obtain a liver tumor segmentation model and a liver vascular segmentation model, respectively. 
Then, we use our liver tumor data and liver vascular data  to retrain the two models obtained by pre-training, and obtain two liver tumor segmentation models and liver vascular segmentation models based on the training sample distribution of our dataset. 
When we need to perform segmentation of liver tumor and blood vessels of CT images, we only need to input one CT image, and the two models will segment the tumor and blood vessel parts of CT images respectively. 
The mask output from the two model segments is automatically fused into a single mask that contains the tumor and blood vessels with relative positions as the final output.

\section{Experiment and discussion}
\label{sec:others}

\subsection{Experimental setup}
\begin{figure}[!t]
	\centering
	\centerline{\includegraphics[width=\columnwidth]{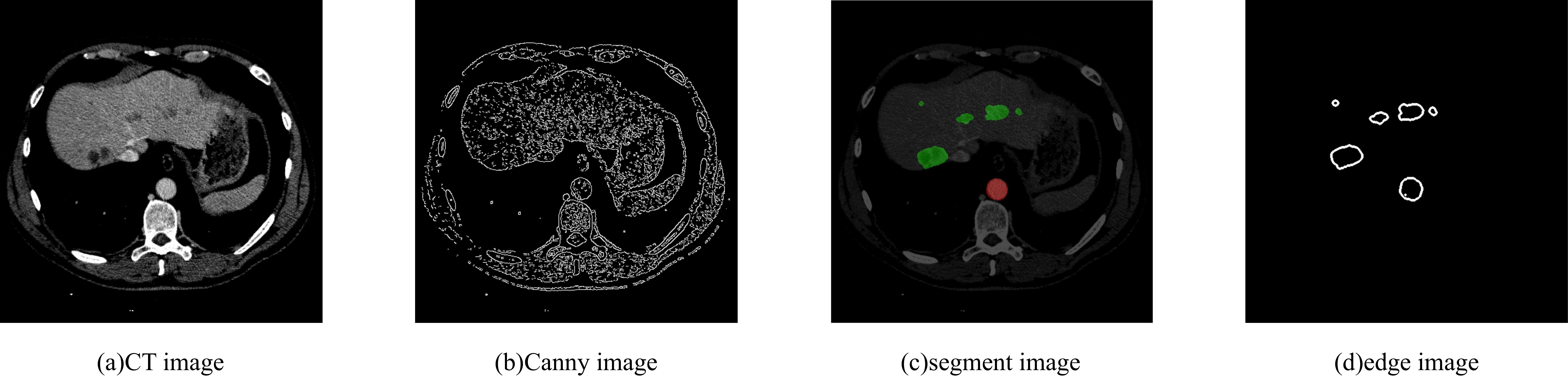}}
	\caption{An example of training dataset (a) input CT image. (b) Image canny map extracted from CT image by Canny algorithm. (c) Segmentation labels of tumors and blood vessels. (d) Edge label for tumors and blood vessels.}
	\label{fig:dataset}
\end{figure}

\subsubsection{Dataset}
The LITS (Liver and Liver Tumor Segmentation, https://competitions.codalab.org/competitions/17094) dataset contains 130 cases of tumors, metastases, and cysts, and these CT scans have large spatial resolution and field of view (FOV) differences\cite{jiang2019ahcnet}.  
3Dircadb (3D Image Reconstruction for Comparison of Algorithm Database, https://www.ircad.fr/research/3d-ircadb-01/) is a public dataset that can be used to train and test liver vessel segmentation methods, including 20 patients in different image resolutions, vessel structure, intensity distribution and liver vessel comparison CT enhancement\cite{huang2018robust}.  
At the same time, we collected 18 typical patients' portal and arterial phase CT enhanced images of the liver for manual annotation, and finally constructed a liver tumor blood vessel(LTBV) dataset. 
The data we used was ethically reviewed, but we cannot make this dataset publicly available because of patient privacy.
We annotated the hepatic arterial vessels in the arterial phase images of the same patient and annotated the liver tumors in the portal phase images. 
Due to the different characteristics of CT images in the two phases, we need to train two models for automatic segmentation of arterial vessels and tumors.

The LITS and 3Dircadb datasets cover a wide range of CT images with different resolution differences and field of view (FOV) differences.
We use these two datasets for model pre-training. We use our private dataset for the fine-tuning training of the model for hepatic artery and tumor segmentation tasks. The above three datasets are divided into training set and test set according to the ratio of 8:2.

\subsubsection{Evaluation Metrics}

In order to better evaluate our model from multiple perspectives, we have selected 5 evaluation indicators including: iou, DSC coefficient, voe, recall, precision.
IoU(Intersection over Union) is the calculation of the intersection of the real annotation and the segmentation result. 
The calculation method was
\begin{equation}
    IoU=\frac{R_{pre}\cap R_{real}}{R_{pre}\cup R_{real}},
    \label{eq6}
\end{equation}
where $R_{pre}$ represents the segmentation result predicted by the model, and $R_{real}$ represents the actual segmentation result.
The DSC(Dice Similariy Coefficient) represents the ratio of the area where the segmented image and the real image intersect to the total area. 
The calculation method was 
\begin{equation}
    DSC=\frac{2\times(R_{pre}\cap R_{real})}{R_{pre} + R_{real}},
    \label{eq7}
\end{equation}
where $R_{pre}$ represents the segmentation result predicted by the model, and $R_{real}$ represents the actual segmentation result.
VOE(Volumetric Overlap Error) represents the difference between the area of the segmented image and the real image, and usually represents the error rate of segmentation. 
The specific calculation method was
\begin{equation}
    VOE=\frac{2\times(R_{pre}-R_{real})}{R_{pre}+R_{real}}.
\end{equation}
Precision is the proportion of pixels that are actually not in the region of interest correctly judged as not in the region of interest. 
It measures the ability to correctly judge the pixels that are not in the region of interest in the segmentation experiment.
Its calculation method was
\begin{equation}
    Precision=\frac{I-R_{pre}\cup R_{real}}{I-R_{real}}.
\end{equation}
Where I is the original input image.
Recall is the proportion of pixels that  are correctly judged as pixels in the region of interest. 
It measures the ability to correctly segment the region of interest.Its calculation method was
\begin{equation}
    Recall=\frac{R_{pre} \cap R_{real}}{R_{real}}.
\end{equation}

\subsection{Performance comparison with state-of-the-art methods}
We choose 5 advanced segmentation models to compare with our method, the 5 models are SegNet\cite{badrinarayanan2017segnet}, UNet\cite{ronneberger2015u}, UNet++\cite{zhou2018unet++}, UNet3+\cite{huang2020unet}, TransUNet\cite{chen2021transunet}. 
At the same time, we divide the proposed method into TransFusionNet(TFN) and TransFusionNet with edge module(TFNEdge), and evaluate them respectively.
We first compare the segmentation effects of five models on blood vessels and tumors based on two public datasets:LITS(Tumor)and 3Dircadb(Vessel). 
Next, we use the LTBV dataset to fine-tune the five models and compare the segmentation effects of the five models.
\subsubsection{Comparison experiment of liver tumor and blood vessel segmentation effects based on public datasets:LITS and 3Dircadb}\label{section}
The performance of TransFusionNet and the other four methods on two public datasets is shown in Table \ref{tab:my_label1}. 
The experimental results show that the IoU of TransFusionNet on the 3Dircadb dataset can reach 0.854, and the DSC can reach 0.918, which is 0.8\% and 1.1\% higher than the IoU and DSC of the baseline method UNet. 
The IoU is 2.3\% and 0.7\% higher than UNet++ and TransUNet, respectively.
However the IoU can reach 0.863 when using TransFusionNet with edge extraction module.
On the LITS dataset, that is, when performing liver tumor segmentation, the IoU and DSC of TransFusionNet can reach 0.840 and 0.910. 
As can be seen from Table \ref{tab:my_label1}, the VOE of TransFusionNet on the two datasets, that is, the error rate is also far lower For other models.


\begin{table}[]
\caption{Performance comparison on LITS and 3Dircadb datasets}
    \centering
    \setlength{\tabcolsep}{3pt}
    \begin{tabular}{lllllll}
    \hline
         Dataset &  Methods & IoU & DSC & VOE & Precision & Recall\\
         \hline
         \multirow{5}{*}{3Dircadb} 
         & SegNet\cite{badrinarayanan2017segnet} & 0.839 & 0.907 & -0.067 & 0.938 & 0.879 \\
         & UNet\cite{ronneberger2015u} & 0.846 & 0.913 & -0.079 & \textbf{0.951} & 0.880 \\
         & UNet++\cite{zhou2018unet++} & 0.831 & 0.904 & -0.062 & 0.934 & 0.879 \\
         & UNet3+\cite{huang2020unet} & 0.853 & 0.917 & -0.059 & 0.945 & 0.894 \\
         & TransUNet\cite{chen2021transunet} & 0.847 & 0.913 & -0.066 & 0.944 & 0.885 \\
         & \textbf{Ours} & 0.854 & 0.918 & -0.041 & 0.938 & 0.901 \\
         & \textbf{Ours with edge module} & \textbf{0.863} & \textbf{0.921} & \textbf{-0.051} & 0.947 & \textbf{0.901} \\
         \hline
         \multirow{5}{*}{LITS}
         & SegNet\cite{badrinarayanan2017segnet} & 0.805 & 0.887 & -0.035 & 0.904 & 0.875 \\
         & UNet\cite{ronneberger2015u} & 0.832 & 0.905 & -0.024 & 0.917 & 0.897 \\
         & UNet++\cite{zhou2018unet++} & 0.828 & 0.902 & -0.020 & 0.912 & 0.896 \\
         & UNet3+\cite{huang2020unet} & 0.821 & 0.895 & -0.002 & 0.899 & 0.898 \\
         & TransUNet\cite{chen2021transunet} & 0.834 & 0.905 & -0.040 & \textbf{0.923} & 0.889 \\
         & \textbf{Ours} & \textbf{0.840} & \textbf{0.910} & \textbf{-0.018} & 0.919 & \textbf{0.904} \\
          & \textbf{Ours with edge module} & \textbf{0.840} & \textbf{0.910} & \textbf{-0.018} & 0.919 & \textbf{0.904} \\
         \hline
    \end{tabular}
    \label{tab:my_label1}
\end{table}

\subsubsection{Comparison experiment of liver tumor and blood vessel segmentation effects based on private dataset:LTBV}

As the experiments described in section \ref{section}, we used LITS and 3Dircadb public datasets to train to obtain segmentation models of liver tumors and blood vessels. 
Compared with the other four state-of-the-art methods, our method has the best automatic segmentation effect. 
We use five methods trained on two datasets to perform transfer learning fine-tuning training on the LTBV dataset.
The performance of each model in the LTBV dataset as shown in Table \ref{tab:my_label2}. 
From Table \ref{tab:my_label2}, we can see that the IoU of TransFusionNet on the blood vessel dataset can reach 0.822, and the DSC can reach 0.899. 
This is 1.9\% and 1.8\% higher than the IoU and DSC of the baseline method SegNet, and is higher than the IoU and Voe of TransUNet. 
They are 0.4\% and 0.5\% higher respectively. On the tumor dataset, the IoU and DSC of TransFusionNet are as high as 0.927 and 0.961, which shows our method can still achieve the best results after LTBV transfer learning.

\begin{table}
\caption{Performance comparison on LTBV datasets}
    \centering
    \setlength{\tabcolsep}{3pt}
    \begin{tabular}{lllllll}
    \hline
         Dataset &  Methods & IoU & DSC & VOE & Precision & Recall\\
         \hline
         \multirow{5}{*}{Vessel} 
         & SegNet\cite{badrinarayanan2017segnet} & 0.803 & 0.881 & -0.056 & 0.907 & 0.858 \\
         & UNet\cite{ronneberger2015u} & 0.812 & 0.893 & \textbf{-0.013} & 0.902 & 0.890 \\
         & UNet++\cite{zhou2018unet++} & 0.809 & 0.892  & -0.058 & 0.919 & 0.868 \\
         & UNet3+\cite{huang2020unet} & 0.821 & 0.897 & 0.022 & 0.892 & \textbf{0.909} \\
         & TransUNet\cite{chen2021transunet} & 0.818 & 0.897 & -0.049 & 0.920 & 0.876 \\
         & \textbf{Ours} & 0.822 & 0.899 & -0.054 & 0.925 & 0.877 \\
         & \textbf{Ours with edge module} & \textbf{0.854} & \textbf{0.901} & -0.040 & \textbf{0.933} & 0.895 \\
         \hline
         \multirow{5}{*}{Tumor}
         & SegNet\cite{badrinarayanan2017segnet} & 0.905 & 0.948 & 0.002 & 0.922 & 0.931 \\
         & UNet\cite{ronneberger2015u} & 0.915 & 0.954 & -0.018 & 0.963 & 0.946 \\
         & UNet++\cite{zhou2018unet++} & 0.912 & 0.952 & 0.003 & 0.952 & 0.954 \\
         & UNet3+\cite{huang2020unet} & 0.827 & 0.899 & -0.037 & 0.918 & 0.885 \\
         & TransUNet\cite{chen2021transunet} & 0.920 & 0.955 & -0.023 & 0.967 & 0.945 \\
         & \textbf{Ours} & \textbf{0.927} & \textbf{0.961} & \textbf{-0.011} & 0.966 & \textbf{0.955} \\
         & \textbf{Ours with edge module} & 0.917 & 0.954 & -0.022 & \textbf{0.975} & 0.945 \\
         \hline
    \end{tabular}
    \label{tab:my_label2}
\end{table}

\subsection{Ablation Study for TransFusionNet model}

\subsubsection{Ablation study of Transformer-based feature extraction module and SEBottleNet local encoder}
\begin{figure}[!t]
	\centering
	\centerline{\includegraphics[width=\columnwidth]{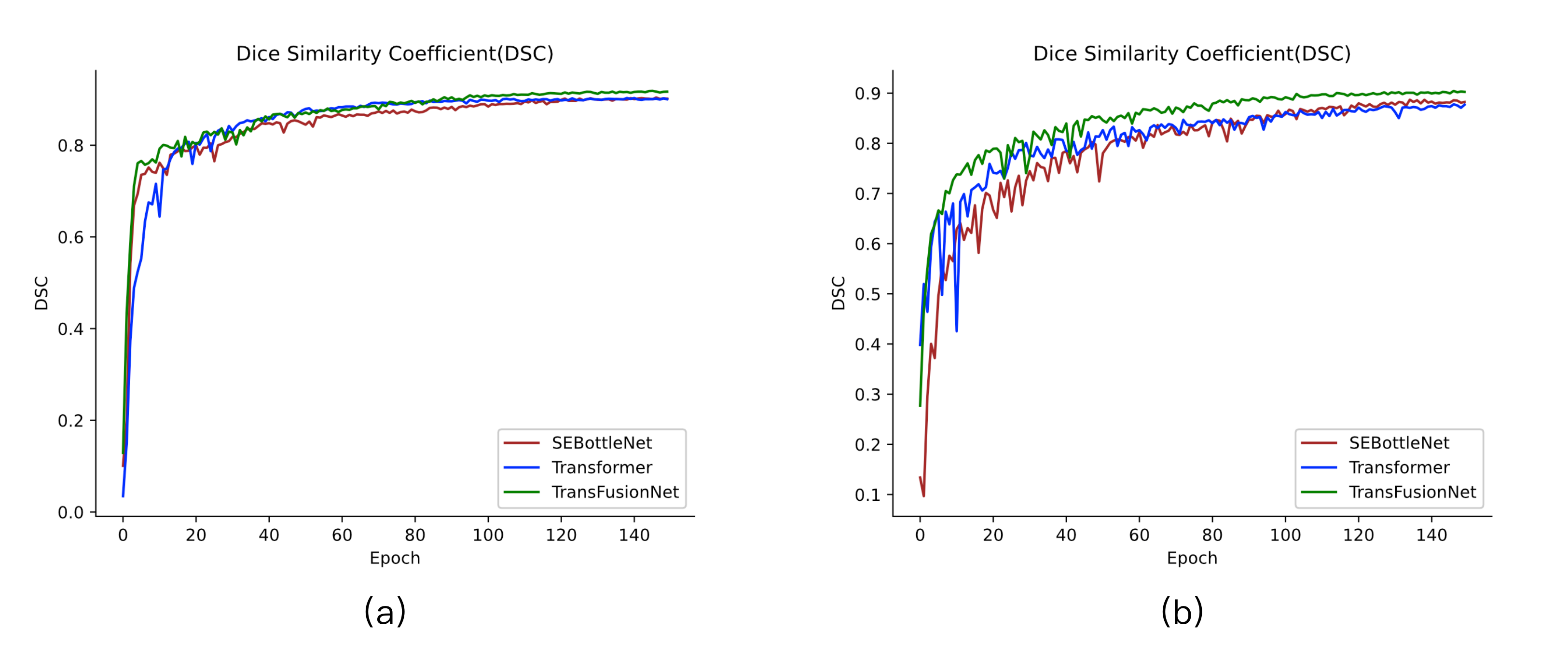}}
	\caption{Ablation experimental results for different feature extraction structures. (a) DSC on 3Dircadb dataset. (b)DSC on LITS dataset.}
	\label{fig:fig5}
\end{figure}

In this section, we use the Transformer module, the Multi-layer local feature extraction module, and TransFusionNet for our experiments, with the aim of testing the effect of the above two modules on the segmentation accuracy of TransFusionNet.
From Figure \ref{fig:fig5}(a) we can see that the Transformer module performs better than the SEBottleNet module in general on the vascular dataset of 3Dircadb.
We believe that the Transformer encoder can learn the CT image global contextual feature representation, especially it encodes the image location information, which certainly helps to enhance the segmentation of the image as a whole.
On the LITS tumor dataset, as shown in Figure \ref{fig:fig5}(b), the segmentation accuracy of the SEBottleNet module is higher, which is attributed to the fact that its internal CNN and local residuals are more interested in some finer features in the image, such as tumor edge features. 
From Figure \ref{fig:fig5}, we can see that this paper achieves an effective improvement in the segmentation accuracy of liver vessels and tumors by combining the Transformer module and SEBottleNet module.

\subsubsection{Function of skip connection used in Decooder}
From results in Tables \ref{tab:my_label1} and \ref{tab:my_label2}, we can find that the TFNEdge module shows the best effect in vascular segmentation task, but the outcome in tumor segmentation task is not as good as TFN. 
Anyway, the tumor segmentation effect of TFNEdge module also exceeds the state-of-the-art model. 
This shows that the edge extraction module can play a good role in the task of small target segmentation.
However, when segmenting large targets, due to the function of the edge loss function, the network will pay more attention to edge optimization and affect the global control of the whole target.

\subsubsection{Function of skip connection used in Decooder}
In the Encoder-Decoder structure, the encoder learns to extract the high-frequency image representation of the feature map, and the decoder continuously learns feature recovery based only on the high-frequency feature coding output from the encoder. 
The role of low-frequency feature information is ignored in the process of encoding and decoding, yet low-frequency features often have their non-negligible role. 
The role of skip connection is to allow the network to better learn low-frequency features during the encoding and decoding process. 
In this experiment, to demonstrate the importance of different modular jump links of our designed network on the segmentation effect, we use TFN Architecture and remove the skip connections of the Transformer module, the skip connections of the local feature extraction module and all skip connections, and train these three models using the same parameter settings. 
The performance gap with the original network is compared. The experimental results are shown in Table \ref{tab:my_label3}.
According to the results in the Table \ref{tab:my_label3}, we can find that the model retaining the global local skip connections has a significant improvement compared to the model with the jump links removed. 
This result proves the importance of skip connections for TransFusionNet and also shows that the low frequency features of the image have a significant impact on the segmentation results.

\begin{table*}
\caption{Performance comparison on LITS and 3Dircadb datasets}
    \centering
    \setlength{\tabcolsep}{3pt}
    \begin{tabular}{lllllll}
    \hline
         Dataset &  Modules without skip connections & IoU & DSC & VOE & Precision & Recall\\
         \hline
         \multirow{4}{*}{3Dircadb} 
         & all encoders & 0.654 & 0.780 & -0.163 & 0.862 & 0.732 \\
         & Transformer-based encoder & 0.767 & 0.858 & 0.059 & 1.066 & 0.886 \\
         & CNN-based encoder & 0.785 & 0.870 & 0.019 & 0.880 & 0.868 \\
         & \textbf{Ours} & \textbf{0.854} & \textbf{0.918} & \textbf{-0.041} & \textbf{0.938} & \textbf{0.901} \\
         \hline
         \multirow{5}{*}{LITS}
         & all encoders & 0.805 & 0.887 & -0.035 & 0.904 & 0.875 \\
         & Transformer-based encoder & 0.832 & 0.905 & -0.024 & 0.917 & 0.897 \\
         & CNN-based encoder & 0.828 & 0.902 & -0.020 & 0.912 & 0.896 \\
         & \textbf{Ours} & \textbf{0.840} & \textbf{0.910} & \textbf{-0.018} & \textbf{0.919} & \textbf{0.904} \\
         \hline
    \end{tabular}
    \label{tab:my_label3}
\end{table*}

\subsection{Visualizations}
\begin{figure*}[!t]
    \centering
    \centerline{\includegraphics[width=\textwidth]{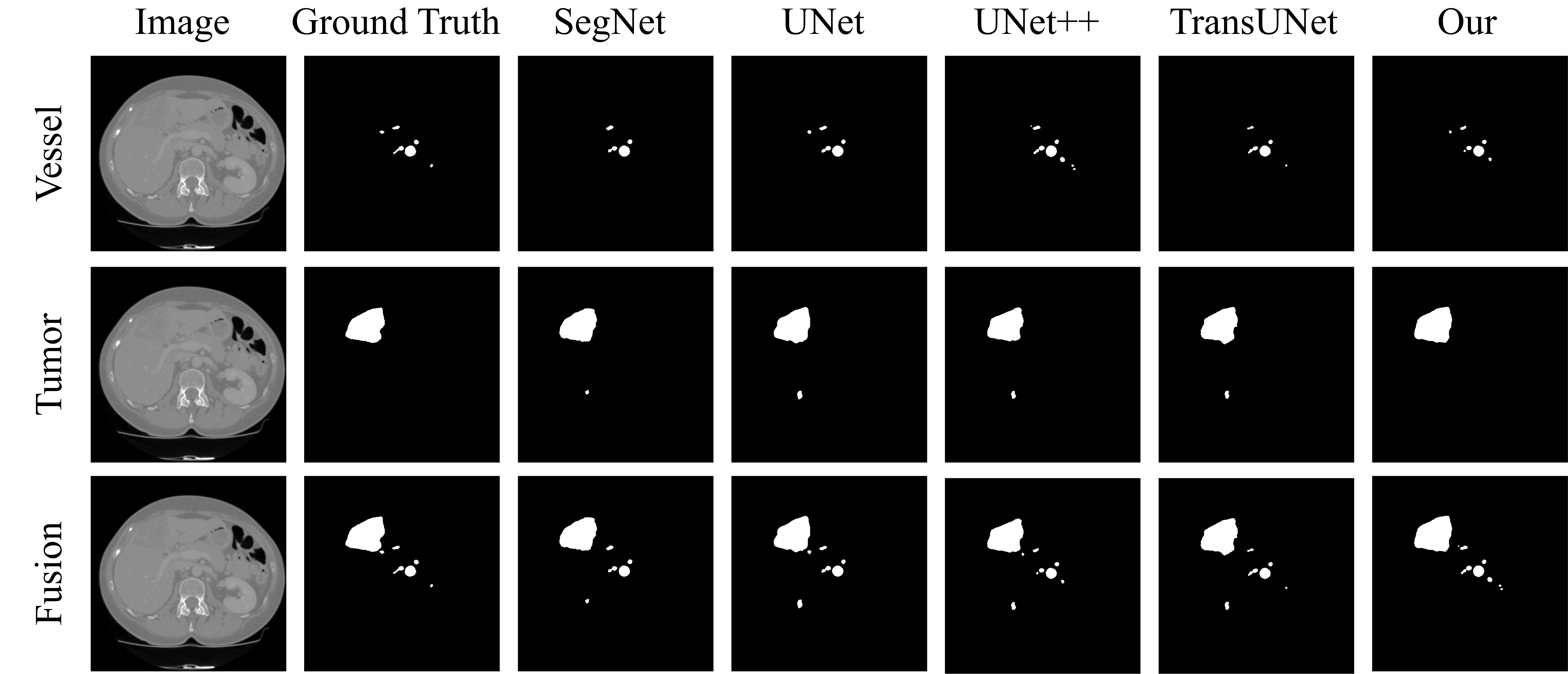}}
    \caption{Qualitative analysis of the 2D segmentation results of blood vessels, tumors and fusions of TransFusionNet and other comparison models from visual perspective}
    \label{fig:fig4}
\end{figure*}

From the above quantitative experimental methods, our model has the best performance in the segmentation of liver blood vessels and liver tumors. 
Next, we use TransFusionNet and other comparable models on a test case on the LTBV dataset to segment liver tumors and blood vessels and then visualize them. 
The first row of Figure \ref{fig:fig4} is the model's segmentation of Vessel, the second row is the model's segmentation of Tumor, and the third row is the result of fusing the first two rows of segmentation results in the same coordinate system.

From the perspective of visual analysis, SegNet and UNet are not accurate in segmenting the details of blood vessels. 
Although UNet++ can identify some details of blood vessels, the error rate is too high. 
TransFusionNet almost perfectly segmented the details of blood vessels, which is more accurate than TransUNet. 
This is attributed to SEBottleNet's extraction of local receptive field information and the importance of different channels. 
In tumor segmentation, UNet++ has a great segmentation result for the edge of the tumor, while SegNet and UNet perform poorly in this respect. 
All comparison models have some wrong tumor segmentation, and TransFusionNet not only avoids these wrong segmentation but also can segment the edge and contour of the tumor accurately. 
We believe that after TransFusionNet extracts global and local information, the multi-scale feature fusion decoder almost perfectly restores the feature of the image, so that the segmentation accuracy is significantly improved, and the error rate is low.

In summary, the above comparison models are not accurate in segmenting tumors and blood vessels. 
They are easy to misclassify some areas that are not tumors, and they are not sensitive to the recognition of some fine blood vessels areas, resulting in incomplete blood vessel segmentation results. 
TransFusionNet can accurately segment the liver tumor regardless of its integrity or vascular continuity.



\subsection{Case Study: 3D Reconstruction of Liver Tumor Vessels Using TransfusionNet under embedded devices}

Medical image segmentation has a wide range of applications and research values in medical research and practice fields such as medical research, clinical diagnosis, pathological analysis, computer-assisted surgery, and three-dimensional simulations. 
In this experiment, we use knowledge distillation to generate a more lightweight model from the TransFusionNet and deploy it into JetsonTX2 embedded device. 
We fed the CT images to the JetsonTX2 embedded system to directly predict the results of 3D reconstruction of liver vessels and tumors.


We define segmentation model $\mathcal{F}:\mathbb{R}^{N \times H \times W}  \rightarrow \mathbb{R}^{N \times C \times H \times W}$. this model has been fully trained. 
Next, the arterial phase CT-enhanced image $x \in \mathbb{R}^{N  \times H \times W}$ feed into $\mathcal{F}$ to predict the liver and vessel label map $o \in \mathbb{R}^{N \times C \times H \times W}$ where $C=3$ in this segmentation task.
We need to construct 3D reconstruction result according to the segmentation label map $o$. Thus, to obtain segmentation result $y \in \mathbb{R}^{N \times H \times W} $ from label map $o$, we use the following equation:
\begin{equation}
    y=\mathcal{G} * argmax(o).
    \label{rec_eq}
\end{equation}
Where $\mathcal{G}$ denotes Gaussian filter.
In order to better support the embedded platform, we refer to the method in \cite{touvron2021training} to distill the knowledge of the Transformer module. 
At the same time, we do the post-processing quantization operation for the middle layer of the trained model.
After the above optimizations, we successfully deployed the model to the embedded processor

\begin{figure}[!t]
    \centering
    \centerline{\includegraphics[width=\columnwidth]{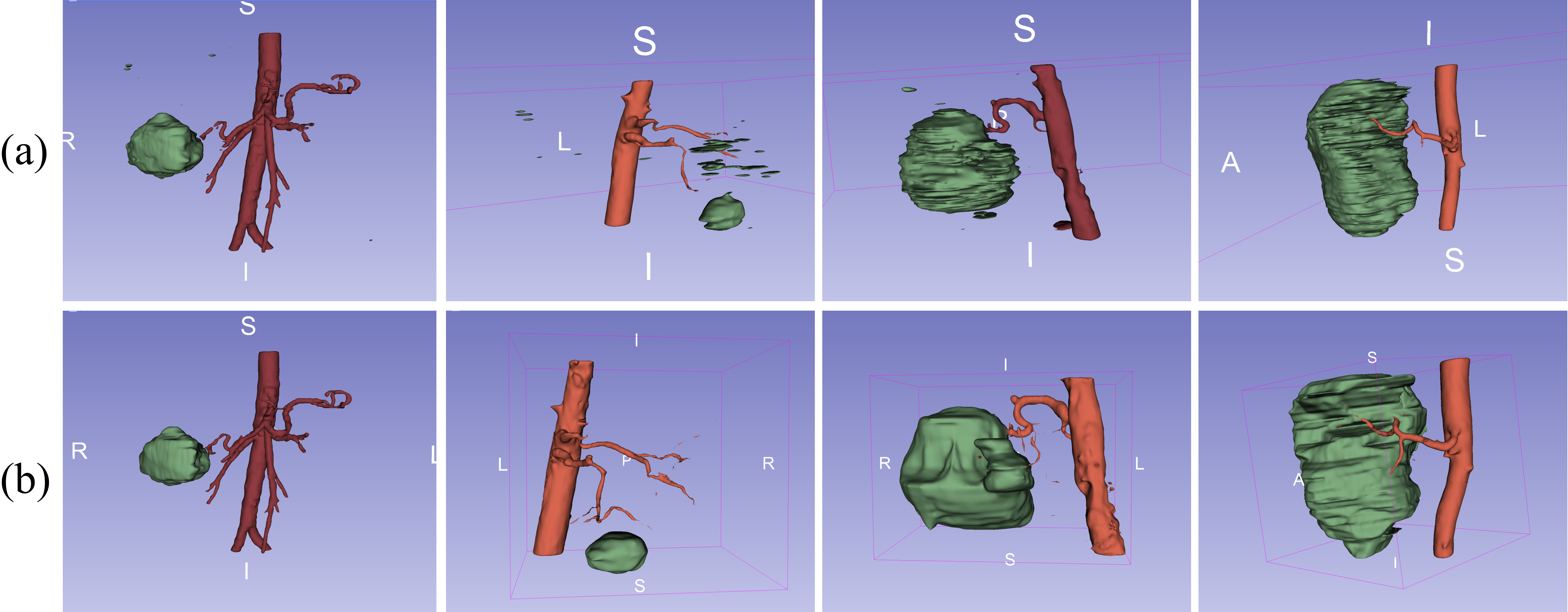}}
    \caption{Visual comparison of reconstruction results. (a)Reconstruction result using automatic reconstruction algorithm under embedded system.(b)Reconstruction results using manual annotations}
    \label{fig:rec}
\end{figure}

Finally, we save the reconstruction results in $.nrrd$ format and use 3Dslicer for visual display.
The Figure \ref{fig:rec} shows a comparison between the reconstruction results under embedded
microprocessor and manual annotations of a typical patient. 
Except for some noise and loss of vessel details, the reconstruction results were very close to the actual annotation results. 
However, a detailed manual annotation requires a lot of time and effort, which significantly reflect the efficiency and accuracy of our proposed algorithm.

\section{Conclusion}
In this work, we propose a segmentation model which can effectively extract the full-scale feature information of CT images. 
The IOU reached the peak performance of 0.864 in the vessel segmentation of the public dataset 3dircadb and 0.840 in the liver tumor segmentation of the public dataset LITS. 
At the same time, we transferred the trained model to our annotated dataset, and the IOU in tumor and vascular segmentation reached 0.927 and 0.822 respectively. 
Compared with the state-of-the-art segmentation methods, TransFusionNet has an accuracy improvement of 1\% - 2\%. 
Although this experiment is only for the segmentation of liver tumors and blood vessels, our model can also be applied to the segmentation of other tissues.

We further compress and distill the model and deploy it to the embedded system. 
Finally, we developed an automatic tumor vessel segmentation and reconstruction device for CT images. And the device realizes the automatic reconstruction of tumors and blood vessels without losing too much accuracy. 
This shows that our method has great application prospects in intelligent surgery in the future.

Although we have completed the reconstruction of liver tumors and blood vessels on Jetson, the segmentation and reconstruction of intrahepatic blood vessels using our method still needs to be further improved.  
Due to the numerous and small branches of intrahepatic vessels, it is difficult for deep learning algorithm to perceive the characteristics of intrahepatic vessels. 
At the same time, the clarity of CT image and the computing power will also hinder the fine reconstruction of internal hepatic artery. 
Facing so many challenges, in the next work, we will further design a novel quantitative method to optimize the inference accuracy of the model in the embedded structure.

\bibliographystyle{IEEEtran}
\bibliography{IEEEabrv, template}

\end{document}